\documentclass[conference]{IEEEtran}
\IEEEoverridecommandlockouts

\usepackage[english]{babel}
\usepackage[pdftex]{graphicx}
\graphicspath{{Figure/}}
\usepackage[cmex10]{amsmath}
\usepackage{amsthm}
\usepackage{amssymb}
\usepackage{array}
\usepackage{color}
\usepackage{epstopdf}
\usepackage{amsfonts}
\usepackage[acronym,shortcuts]{glossaries}
\usepackage{nccmath}

\usepackage{threeparttable}
\usepackage{booktabs}
\usepackage{cuted}
\usepackage{siunitx}
\DeclareSIUnit{\Wh}{Wh}
\DeclareSIUnit{\belmilliwatt}{Bm}
\DeclareSIUnit{\dBm}{\deci\belmilliwatt}
\DeclareSIUnit{\belmillii}{Bi}
\DeclareSIUnit{\dBi}{\deci\belmillii}

\usepackage[font=footnotesize]{caption}
\usepackage{subcaption}

\usepackage{algorithm}
\usepackage{algpseudocode}
\usepackage{float}

\usepackage[dvipsnames]{xcolor}

\usepackage{pgfplots}
\pgfplotsset{
    compat=1.14,
    grid style={black!60!white, thin, densely dotted},
    every axis plot/.append style={no markers, thick},
    label style={font=\footnotesize},
    tick label style={font=\footnotesize}
}
\usepackage{tikz}							% Tikz
\usetikzlibrary{shapes,arrows,fit,positioning}
\usetikzlibrary{shapes}
\usetikzlibrary{spy}
\usetikzlibrary{circuits}
\usetikzlibrary{arrows}

\newcommand{\vect}[1]{\boldsymbol{\mathrm{#1}}}
\newcommand{\mat}[1]{\boldsymbol{\mathrm{#1}}}

\newcommand{\tr}{\mathrm{tr}}

\newcommand{\diag}{\mathrm{diag}}

\newcommand{\update}[1]{#1}

\newcommand{\abs}[1]{\left\lvert#1\right\rvert}

\newcommand{\cn}[2]{\ensuremath{\sim\mathcal{C}\mathcal{N}\left(#1,#2\right)}}
\newcommand{\norm}[1]{\left\lVert#1\right\rVert}
\newcommand{\normalized}[1]{\bar{#1}}
\newcommand*{\hermconj}[1]{{#1}^{H}}

\usepackage[backend=biber, style=ieee, citestyle=numeric-comp,url=true,eprint=true,doi=true,isbn=true,maxnames=99]{biblatex}
\AtBeginBibliography{\footnotesize}

\usepackage{mathtools}

\theoremstyle{definition}

\makeglossaries
\newacronym{3gpp}{3GPP}{3rd Generation Partnership Project}
\newacronym{abp}{ABP}{Authentication By Personalisation}
\newacronym{aclr}{ACLR}{adjacent channel leakage ratio}
\newacronym{adr}{ADR}{Adaptive Data Rate}
\newacronym{ag}{AG}{array gain}
\newacronym{ai}{AI}{Artificial Intelligence}
\newacronym{amam}{AM/AM}{amplitude modulation to amplitude modulation}
\newacronym{amp}{AMP}{approximate message passing}
\newacronym{ampm}{AM/PM}{amplitude modulation to phase modulation}
\newacronym{aoa}{AOA}{angle-of-arrival}
\newacronym{aod}{AOD}{angle-of-departure}
\newacronym{ap}{AP}{access point} 
\newacronym{ar}{AR}{augmented reality}
\newacronym{arp}{ARP}{Antenna Reference Point}
\newacronym{asic}{ASIC}{application specific integrated circuit}
\newacronym{auv}{AUV}{autonomous underwater vehicle}
\newacronym{awgn}{AWGN}{additive white Gaussian noise}
\newacronym{bom}{BOM}{bill of materials}
\newacronym[plural=BSs,firstplural=Base Stations (BSs)]{bs}{BS}{base station}
\newacronym{bldc}{BLDC}{brushless DC}
\newacronym{bw}{BW}{Bandwidth}
\newacronym{cbm}{CBM}{Condition Based Maintenance}
\newacronym{cc}{CC}{constant current}
\newacronym{ccdf}{CCDF}{complementary cumulative distribution function}
\newacronym{ccnn}{CCNN}{circular convolutional neural network}
\newacronym{cdf}{CDF}{cumulative distribution function}
\newacronym{cdrx}{CDRX}{Connected Mode DRX}
\newacronym{ce}{CE}{coverage enhancement}
\newacronym{ced}{CED}{cumulative energy density}
\newacronym{cf}{CF}{cell-free}
\newacronym{cfo}{CFO}{carrier frequency offset}
\newacronym{cots}{COTS}{commercial off-the-shelf}
\newacronym{cp}{CP}{cyclic prefix}
\newacronym{cpt}{CPT}{capacitive power transfer}
\newacronym{cpu}{CPU}{central-processing unit}
\newacronym{cqi}{CQI}{channel quality indicator}
\newacronym{cr}{CR}{Coding Rate}
\newacronym{crlb}{CRLB}{Cram\'er-Rao lower bound}
\newacronym{crs}{CRS}{cell reference signal}
\newacronym{cs}{CS}{compressed sensing}
\newacronym{csi}{CSI}{channel state information}
\newacronym{csp}{CSP}{contact service point}
\newacronym{css}{CSS}{Chirp Spread Spectrum}
\newacronym{cv}{CV}{constant voltage}
\newacronym{dc}{DC}{direct current}
\newacronym{dcc}{DCC}{dynamic cooperation clustering}
\newacronym{de}{DE}{drain efficiency}
\newacronym{dl}{DL}{downlink}
\newacronym{dmimo}{D-MIMO}{distributed MIMO}
\newacronym{dpd}{DPD}{digital pre-distortion}
\newacronym{drx}{DRX}{Discontinuous Reception Mode}
\newacronym{e2e}{E2E}{End-to-End}
\newacronym{easa}{EASA}{European Union Aviation Safety Agency}
\newacronym{ecdf}{eCDF}{empirical cumulative distribution function}
\newacronym{ecsp}{ECSP}{edge computing service point}
\newacronym{edlc}{EDLC}{electrostatic double-layer capacitors}
\newacronym{edrx}{eDRX}{Extended Discontinuous Reception Mode}
\newacronym{ee}{EE}{energy efficiency}
\newacronym{egprs}{EGPRS}{Enhanced Data Rates for GSM Evolution}
\newacronym{eirp}{EIRP}{effective isotropic radiated power}
\newacronym{em}{EM}{electromagnetic}
\newacronym{embb}{eMBB}{enhanced Mobile Broadband}
\newacronym{en}{EN}{energy neutral}
\newacronym{eol}{EoL}{End of Life}
\newacronym{epu}{EPU}{edge processing unit}
\newacronym{erp}{ERP}{effective radiated power}
\newacronym[plural=ESCs,firstplural=electronic speed controllers (ESCs)]{esc}{ESC}{electronic speed control}
\newacronym{etsi}{ETSI}{European Telecommunications Standards Institute}
\newacronym{evm}{EVM}{Error Vector Magnitude}
\newacronym{fa}{FA}{federation anchor}
\newacronym{fft}{FFT}{fast Fourier transform}
\newacronym{fh}{FH}{fronthaul}
\newacronym{fim}{FIM}{Fisher information matrix}
\newacronym{fom}{FoM}{figure of merit}
\newacronym{gb}{GB}{grant-based}
\newacronym{gf}{GF}{grant-free}
\newacronym{gnb}{gNB}{Next Generation Node B}
\newacronym{gnn}{GNN}{graph neural network}
\newacronym{gnss}{GNSS}{Global Navigation Satellite System}
\newacronym{gprs}{GPRS}{General Packet Radio Services}
\newacronym{gps}{GPS}{Global Positioning System}
\newacronym{gpu}{GPU}{graphics processing unit}
\newacronym{gsm}{GSM}{Global System for Mobile Communications}
\newacronym{gwp}{GWP}{Global Warming Potential}
\newacronym{harq}{HARQ}{hybrid automatic repeat request}
\newacronym{hcs}{HCS}{human-centric services}
\newacronym{hpbm}{HPBM}{half power beam width}
\newacronym{HyMPRo}{HyMPRo}{Hybrid Multi-Path Routing algorithm}
\newacronym{ib}{IB}{in-band}
\newacronym{ibo}{IBO}{input back-off}
\newacronym{ic}{IC}{integrated circuit}
\newacronym{id}{ID}{information decoding}
\newacronym{iid}{i.i.d.}{independently and identically distributed}
\newacronym{im}{IM}{intermodulation}
\newacronym{imu}{IMU}{inertial measurement unit}
\newacronym{iot}{IoT}{Internet of Things}
\newacronym{ipt}{IPT}{inductive power transfer}
\newacronym{ipy}{IPY}{Interventions per Year}
\newacronym{ism}{ISM}{Industrial, Scientific and Medical}
\newacronym{isp}{ISP}{internet service provider}
\newacronym{kpi}{KPI}{key performance indicator}
\newacronym{kvi}{KVI}{key value indicator}
\newacronym{lca}{LCA}{life cycle assessment}
\newacronym{lco}{LCO}{lithium cobalt oxide}
\newacronym{ldo}{LDO}{Low-dropout voltage regulator}
\newacronym{ldpc}{LDPC}{low-density parity-check}
\newacronym{lic}{LIC}{lithium-ion capacitor}
\newacronym{lfp}{LFP}{lithium iron phosphate}
\newacronym{liion}{Li-ion}{lithium-ion}
\newacronym{lipo}{LiPo}{lithium polymer}
\newacronym{lis}{LIS}{large intelligent surface}
\newacronym{llh}{LLH}{log-likelihood}
\newacronym{lmmse}{LMMSE}{least minimum mean square error}
\newacronym{lmo}{LMO}{lithium ion manganese oxide}
\newacronym{lora}{LoRa}{Long Range}
\newacronym{lorawan}{LoRaWAN}{long-range wide-area network}
\newacronym{los}{LoS}{line-of-sight}
\newacronym{lp}{LP}{linear programming}
\newacronym{lpt}{LPT}{laser power transfer}
\newacronym{lpwa}{LPWA}{Low Power Wide Area}
\newacronym{lpwan}{LPWAN}{low-power wide-area network}
\newacronym{lrelu}{LReLU}{leaky rectified linear unit }
\newacronym{lrt}{LRT}{likelihood-ratio test}
\newacronym{ls}{LS}{least squares}
\newacronym{lsfc}{LSFC}{large-scale fading component}
\newacronym{ltc}{LTC}{lithium thionyl chloride}
\newacronym{lte}{LTE}{Long Term Evolution}
\newacronym{lto}{LTO}{lithium titanate}
\newacronym{m2m}{M2M}{Machine to Machine}
\newacronym{mac}{MAC}{Medium Access Control}
\newacronym{mcl}{MCL}{Maximum Coupling Loss}
\newacronym{mcs}{MCS}{modulation and coding scheme}
\newacronym{mcu}{MCU}{Microcontroller Unit}
\newacronym{mec}{MEC}{multi-access edge computing}
\newacronym{mf}{MF}{matched filter}
\newacronym{mmimo}{mMIMO}{massive MIMO}
\newacronym{mimo}{MIMO}{multiple-input multiple-output}
\newacronym{miso}{MISO}{multiple-input single-output}
\newacronym{ml}{ML}{maximum-likelihood}
\newacronym{mlp}{MLP}{multilayer perceptron}
\newacronym{mmse}{MMSE}{minimum mean square error}
\newacronym{mmtc}{mMTC}{massive machine-typed communication}
\newacronym{mmwave}{mmWave}{millimeter wave}
\newacronym{mpc}{MPC}{multipath component}
\newacronym{mppt}{MPPT}{maximum power point tracking}
\newacronym{mr}{MR}{maximum ratio}
\newacronym{mrc}{MRC}{Magnetic Resonance Coupling}
\newacronym{mrc_em}{MRC}{maximum ratio combining}
\newacronym{mrt}{MRT}{maximum ratio transmission}
\newacronym{mse}{MSE}{mean square error}
\newacronym{mtc}{MTC}{Machine-Type Communication}
\newacronym{multi-rat}{Multi-RAT}{Multiple Radio Access Technology}
\newacronym{nb}{NB}{narrowband}
\newacronym{nbiot}{NB-IoT}{Narrowband IoT}
\newacronym{nca}{NCA}{nickel cobalt aluminum}
\newacronym{nicd}{NiCd}{nikkel cadmium}
\newacronym{nimh}{NiMH}{nikkel metal hydride}
\newacronym{nlos}{NLoS}{non-line-of-sight}
\newacronym{nmc}{NMC}{nickel manganese cobalt}
\newacronym{nn}{NN}{neural network}
\newacronym{nnls}{NNLS}{non-negative least squares}
\newacronym{noma}{NOMA}{non-orthogonal multiple access}
\newacronym{nr}{NR}{New Radio}
\newacronym{oai}{OAI}{OpenAirInterface} % no spelling mistake, it is spelt all glued to each other...
\newacronym{ofdm}{OFDM}{orthogonal frequency-division multiplexing}
\newacronym{ofdmim}{OFDM-IM}{OFDM with index modulation}
\newacronym{oob}{OOB}{out-of-band}
\newacronym{ota}{OTA}{over-the-air}
\newacronym{otaa}{OTAA}{Over The Air Authentication}
\newacronym{p2p}{P2P}{point-to-point}
\newacronym[plural=PAs,firstplural=power amplifiers (PAs)]{pa}{PA}{power amplifier}
\newacronym{pae}{PAE}{power-added efficiency}
\newacronym{papr}{PAPR}{peak-to-average power ratio}
\newacronym{pc}{PC}{pilot count}
\newacronym{pcb}{PCB}{Printed Circuit Board}
\newacronym{pdcch}{PDCCH}{physical downlink control channel}
\newacronym{pdp}{PDP}{power delay profile}
\newacronym{pdsch}{PDSCH}{physical downlink shared channel}
\newacronym{per}{PER}{packet error rate}
\newacronym{pg}{PG}{path gain}
\newacronym{pl}{PL}{path loss}
\newacronym{prbs}{PRBs}{Physical Resource Blocks}
\newacronym{psd}{PSD}{power spectral density}
\newacronym{psm}{PSM}{Power Saving Mode}
\newacronym{ptw}{PTW}{Paging Time Window}
\newacronym{pcg}{PCG}{power consumption gain}
\newacronym{pgd}{PGD}{proximal gradient descent}
\newacronym{qam}{QAM}{quadrature amplitude modulation}
\newacronym{qos}{QoS}{quality-of-service}
\newacronym{quadriga}{QuaDRiGa}{QUAsi Deterministic RadIo channel GenerAtor}
\newacronym{ra}{RA}{Random Access}
\newacronym{ran}{RAN}{radio access network}
\newacronym{rar}{RAR}{Random Access Response}
\newacronym{rat}{RAT}{Radio Access Technology}
\newacronym{rbs}{RBS}{radio base station}
\newacronym{re}{RE}{radio element}
\newacronym[]{relu}{ReLU}{rectified linear unit}
\newacronym{rf}{RF}{radio frequency}
\newacronym{rfeh}{RFEH}{radio frequency energy harvesting}
\newacronym{rfid}{RFID}{radio frequency identification}
\newacronym{rfpt}{RFPT}{radio frequency power transfer}
\newacronym{ris}{RIS}{reflective intelligent surface}%reconfigurable?
\newacronym{rms}{RMS}{root-mean-square}
\newacronym{rmse}{RMSE}{root-mean-square error}
\newacronym{rrc}{RRC}{Radio Resource Connection}
\newacronym{rreq}{RREQ}{route request packet}
\newacronym{rsrp}{RSRP}{Reference Signals Received Power}
\newacronym{rss}{RSS}{received signal strength}
\newacronym{rssi}{RSSI}{received signal strength indicator}
\newacronym{rtc}{RTC}{Real Time Clock}
\newacronym{rtk}{RTK}{Real Time Kinematics}
\newacronym{rts}{RTS}{ray tracing simulator}
\newacronym{rw}{RW}{RadioWeaves}
\newacronym{rx}{RX}{receiver}
\newacronym{rzf}{RZF}{regularized zero forcing}
\newacronym{sa}{SA}{synchronization anchor}
\newacronym{scfdma}{SCFDMA}{Single-Carrier Frequency Division Multiple Access}
\newacronym{sdg}{SDG}{Sustainable Development Goal}
\newacronym{sdn}{SDN}{software-defined network}
\newacronym{sdr}{SDR}{signal-to-distortion ratio}
\newacronym{sf}{SF}{Spreading Factor}
\newacronym{sfn}{SFN}{single frequency network}
\newacronym{sinr}{SINR}{signal-to-interference-plus-noise ratio}
\newacronym{siso}{SISO}{single-input single-output}
\newacronym{slam}{SLAM}{simultaneous localization and mapping}
\newacronym{slc}{SLC}{spatial leakage suppression}
\newacronym{smc}{SMC}{specular multipath component}
\newacronym{smps}{SMPS}{Switched Mode Power Supply}
\newacronym{sndr}{SNDR}{signal-to-noise-and-distortion ratio}
\newacronym{snidr}{SNIDR}{signal-to-noise-and-interference-and-distortion ratio}
\newacronym{snr}{SNR}{signal-to-noise ratio}
\newacronym{soc}{SoC}{state of charge}
\newacronym{ssb}{SSB}{synchronisation signal block}
\newacronym{steam}{STEAM}{science, technology, engineering, the arts, and mathematics}
\newacronym{svd}{SVD}{singular value decomposition}
\newacronym{tau}{TAU}{Tracking Area Update}
\newacronym{tdd}{TDD}{time division duplexing}
\newacronym{tdoa}{TDOA}{time-difference-of-arrival}
\newacronym{toa}{TOA}{time-of-arrival}
\newacronym{trp}{TRP}{Transmission Reception Point}
\newacronym{ttm}{TTM}{Time To Market}
\newacronym{ttn}{TTN}{The Things Network}
\newacronym{tx}{TX}{transmitter}
\newacronym{tcer}{TCER}{transported to consumed energy ratio}
\newacronym{uav}{UAV}{unmanned aerial vehicle}
\newacronym{udp}{UDP}{User Datagram Protocol}
\newacronym{ue}{UE}{user equipment}
\newacronym{ugv}{UGV}{unmanned ground vehicle}
\newacronym{uhf}{UHF}{ultra high frequency}
\newacronym{ul}{UL}{uplink}
\newacronym{ula}{ULA}{uniform linear array}
\newacronym{upa}{UPA}{uniform planar array}
\newacronym{ura}{URA}{uniform rectangular array}
\newacronym{urllc}{URLLC}{Ultra-Reliable Low Latency Communications}
\newacronym{uv}{UV}{unmanned vehicle}
\newacronym{uwb}{UWB}{ultrawideband}
\newacronym{wb}{WB}{wideband}
\newacronym{wpt}{WPT}{wireless power transfer}
\newacronym[plural=WRSNs,firstplural=Wireless Rechargeable Sensor Networks (WRSNs)]{wrsn}{WRSN}{Wireless Rechargeable Sensor Network}
\newacronym[plural=WSNs,firstplural=Wireless Sensor Networks (WSNs)]{wsn}{WSN}{Wireless Sensor Network}
\newacronym{z3ro}{Z3RO}{zero third-order distortion}
\newacronym{zf}{ZF}{zero-forcing}
\newacronym{zmcscg}{ZMCSCG}{zero mean circularly symmetric complex Gaussian}

\usepackage{dblfloatfix}

\usepackage{xifthen}
\newcommand{\prob}[1][]{%requires xifthen package%
\ifthenelse{\isempty{#1}}%
      {\ensuremath{P}}%
    {\ensuremath{P\left\(#1\right\)}}%
}
\usepackage[hidelinks]{hyperref}
\usepackage{url}
\usepackage{authblk}
\usepackage{nicefrac}

\pgfplotsset{
    every axis/.append style={
        title style={draw=none},
        label style={font=\footnotesize},
        legend style={
            fill opacity=0.8,
            nodes={scale=0.8, transform shape}, {draw=none}
        },
        tick align=outside,
        tick pos=left,
        x grid style={darkgray176},
        xtick style={color=black},
        y grid style={darkgray176},
        ytick style={color=black},
        grid=both,
    },
    every axis plot/.append style={
        line width=2.0pt,
    },
}

\title{Grant-Free Random Access of IoT devices in Massive MIMO with Partial CSI}

 \author[1]{Gilles Callebaut}
 \author[1]{Fran\c{c}ois Rottenberg}
 \author[1]{Liesbet Van der Perre}
 \author[2]{Erik G. Larsson\thanks{The research reported herein was partly funded by the European Union’s Horizon 2020 research and
innovation programme under grant agreement No 101013425.}\thanks{This paper is presented at IEEE~WCNC 2023. \fullcite{Call2303:Grant}}}

\affil[1]{\small Department of Electrical Engineering (ESAT-DRAMCO), KU Leuven, 9000 Ghent, Belgium}
\affil[2]{\small Department of Electrical Engineering (ISY), Link\"{o}ping University, Link\"{o}ping, Sweden}

\def\BibTeX{{\rm B\kern-.05em{\sc i\kern-.025em b}\kern-.08em
    T\kern-.1667em\lower.7ex\hbox{E}\kern-.125emX}}
    
\begin{document}\sloppy

% \author{Gilles Callebaut,~Fran\c{c}ois Rottenberg,
% Liesbet Van der Perre and~Erik G. Larsson\vspace{-0em}% <-this % stops a space
% \thanks{Gilles Callebaut, Liesbet Van der Perre and Fran\c{c}ois Rottenberg is with ESAT-DRAMCO, Ghent Technology Campus, KU Leuven, 9000 Ghent, Belgium (e-mail: gilles.callebaut@kuleuven.be).}
%  \thanks{Erik G. Larsson is with the Department of Electrical Engineering (ISY), Link\"{o}ping University, Link\"{o}ping, Sweden.}
% 	\thanks{The research reported herein was partly funded by the European Union’s Horizon 2020 research and
% innovation programme under grant agreement No 101013425.} 
	
%}

% \markboth{CONFIDENTIAL}%
% {}
%\markboth{Journal of \LaTeX\ Class Files,~Vol.~13, No.~9, September~2014}%
%{Shell \MakeLowercase{\textit{et al.}}: Bare Demo of IEEEtran.cls for Journals}



\maketitle
\vspace{-1cm}
\begin{abstract}
The number of wireless devices is drastically increasing, resulting in many devices contending for radio resources. %This problem of random access is widely studied, which has yielded a variety of random access protocols. 
In this work, we present an algorithm to detect active devices for unsourced random access, \textit{i.e.,} the devices are uncoordinated. 
The devices use a unique, but non-orthogonal preamble, known to the network, prior to sending the payload data. They do not employ any carrier sensing technique and blindly transmit the preamble and data.  To detect the active users, we exploit partial \gls{csi}, which could have been obtained through a previous channel estimate. For static devices, \textit{e.g.,} \acrlong{iot} nodes, it is shown that \gls{csi} is less time-variant than assumed in many theoretical works. 
The presented iterative algorithm uses a maximum likelihood approach to estimate both the activity and a potential phase offset of each known device. The convergence of the proposed algorithm is evaluated. The performance in terms of probability of miss detection and false alarm is assessed for different qualities of partial \gls{csi} and different \acrlong{snr}.
\end{abstract}

\begin{IEEEkeywords}
activity detection, grant-free, massive MIMO, maximum likelihood, random access.
\end{IEEEkeywords}

% \fi
\IEEEpeerreviewmaketitle

\glsresetall

% \gilles{added toc to give us an overview}
% {\footnotesize\tableofcontents}

\section{Introduction}
\Gls{mmtc} is envisioned to enable low-power connectivity to a very large number of devices and open up new applications. While it has been put forward as a key building block in 5G and beyond, it has so far received less attention than \gls{embb} and \gls{urllc}.
% While current endeavors in 5G and beyond systems have focused on providing high spectral efficiency (\gls{embb}) and low-latency (\gls{urllc}), \gls{mmtc} has received less attention.
The challenge in \gls{mmtc}, and in general \gls{iot} systems, resides in the low-power operation, sporadic nature of the traffic and a large amount of uncoordinated devices. As these devices are often battery-powered, they are constrained in the signalling overhead they can handle~\cite{s21030913}. Furthermore, they are often deployed in remote areas, where network coverage is low or non-existent~\cite{s21030913,9443344}. To address this, novel protocols need to be designed tailored to the low-power, sporadic and massive nature of \gls{mmtc} traffic. One promising direction, and the focus of this work, is the use of massive MIMO, where a large number of antennas is present at the base station.

Two multiple access approaches can be taken, grant-based and grant-free random access. The former requires the devices to obtain a grant from the network, where after it can use  a collision-free radio resource to transmit its data. The disadvantage of this approach for \gls{iot} and \gls{mmtc} is that devices need to compete for grants, requiring, \textit{e.g.,} dedicated preambles or a lot of signalling for collision resolution. Due to the number of competing devices such schemes are not practical and scalable.  

Therefore, grant-free approaches are advocated over grant-based solutions~\cite{8454392, 8456557, 9049039}. In grant-free random access, devices do not contend for a grant, but just access the network when required. Often these devices are unaware of the other devices in the system and operate in an uncoordinated fashion, which is called unsourced grant-free random access~\cite{9049039}. Different approaches have been studied to detect the active devices in massive MIMO systems. Detecting active devices is facilitated when each device uses a unique and orthogonal preamble sequence. This entails that each preamble should have the same length as the number of devices to have no collisions, which is unpractical. Therefore a number of studies have considered a pool of orthogonal pilots~\cite{9091017} or the use of non-orthogonal pilots. Both of these strategies have been elaborated in~\cite{8454392}, where they formulate the device-activity detection problem as a compressed sensing problem, where the sparsity of the active devices at each time slot is exploited. However, compressed sensing requires the preamble length to be larger than the active devices, increasing the energy expenditure of data transfer. To combat this, in~\cite{9049039, FenglerTIT} a covariance-based method is suggested using two estimators for device activity recovery, \textit{i.e.,} \gls{ml} and \gls{nnls}. This work is generalized by \citeauthor{GanesanSPAWC}~\cite{GanesanSPAWC,GanesanTCOM} to the cell-free case where multiple \glspl{ap} are geographically distributed. They considered different large-scale fading coefficients per \glspl{ap}. They employed a simplification of their proposed algorithm by including only the \gls{ap} with the strongest contribution. They concluded that co-located massive MIMO is highly sensitive to low \glspl{snr}, while cell-free deployment is better suited against shadowing fading effects. 

\textbf{Contributions.} 
Inspired by these works, we propose a new algorithm exploiting the static nature of \gls{iot} devices. As demonstrated in~\cite{callebaut2021grant} and elaborated in Secion~\ref{section:motivation}, the \gls{csi} can remain almost invariant over large period of times (hours) for scenarios with low mobility. This was not yet leveraged by practical algorithms in the literature to the best of our knowledge. Given that the base station knows a part of the \gls{csi} of each device the performance of the device activity scheme can be improved. To do so, we formulate the \acrlong{ml} activity detection problem using partial \gls{csi}. %The ``amount of partial CSI'' is a priori known can be adapted for each user through a parameter, which can model all cases going from no prior \gls{csi} to full prior \gls{csi}. 
Furthermore, a phase offset can be estimated which occurs due to \gls{cfo} (when the \gls{cfo} is static over the preamble duration). %Besides presenting an analytical derivation of the proposed algorithm, a numerical evaluation is performed. 
We validate the convergence of the iterative algorithm, study the impact of different initialization vectors for the device activity, the impact of the quality of the partial \gls{csi} and the \gls{snr} on the performance of the algorithm. 

The used mathematical notations are described in \url{https://github.com/wavecore-research/math-notations}.

\section{\update{Motivation -- Recurrence in \gls{csi}}}\label{section:motivation}
\Gls{iot} technologies are often put in the field with a deploy-and-forget strategy, where the devices remain immobile afterwards. 
As such, we can expect that the channel conditions are less time-variant than typically assumed in literature~\cite{SIG-093,marzetta2016fundamentals}.  An experimental campaign to investigate the long-term behavior of the channel is presented in~\cite{callebaut2021grant}. The long-term behavior is measured by taking the channel correlation
$ \delta_{i,j} = 
    \nicefrac
    {\abs{
    \hermconj{\normalized{\vect{h}}_{i}} \cdot \normalized{\vect{h}}_{j}
    }}
    {\norm{\normalized{\vect{h}}_{i}} \norm{\normalized{\vect{h}}_{j}}}$
of the first channel estimate and the other channel $N$ measurements, i.e., $\delta_{1,j}\ \text{for}\ j \in \{1,\dotsc,N\}$.

% \begin{equation}\label{eq:sub-mamimo-corr-coeff}
%     \delta_{i,j} = 
%     \frac
%     {\abs{
%     \hermconj{\normalized{\vect{h}}_{i}} \cdot \normalized{\vect{h}}_{j}
%     }}
%     {\norm{\normalized{\vect{h}}_{i}} \norm{\normalized{\vect{h}}_{j}}}
% \end{equation}

\begin{figure}[hbtp]
    \centering
    \definecolor{darkgray167}{RGB}{0,0,0}
        \input{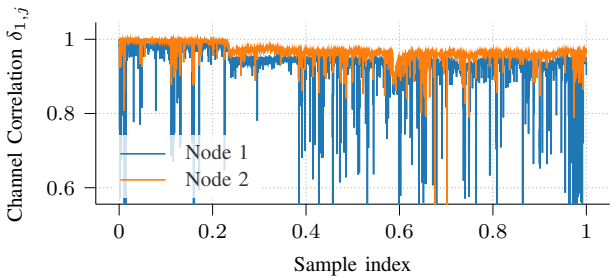}
    \caption{Channel correlation over a full day (9h24-17h48) with over \num{10000} channel instances, with an average correlation of \num{0.935}/\num{0.968} and a standard deviation of \num{0.06}/\num{0.03} for Node~1/Node~2.}\label{fig:init-access-corr-full-day}
\end{figure}

The observed channel correlations are depicted in Fig.~\ref{fig:init-access-corr-full-day}. It illustrates that most of the time the correlation coefficient is close to \num{1}, indicating that the channel is highly correlated with the first estimate and thus can be considered static. It also shows that, while in some occasions the correlation drops, the channel quickly becomes again highly correlated with the first channel instance. More than 90\% of the measured channels\footnote{More specifically, 91\% and 95\% for Node~1 and Node~2, respectively.} have a correlation coefficient higher than \num{0.9}. This demonstrates the potential of re-using channel estimates in \gls{iot} contexts.\footnote{We use here the term ``re-using'' to indicate that we no longer operate in a block fading model with independent channel realizations. Typically, these blocks are considered in the order of \SI{50}{\milli\second}.}

% More than 90\% of the measured channels\footnote{More specifically, 91\% and 95\% for Node~1 and Node~2, respectively.} have a correlation coefficient higher than \num{0.9} over a window of more than \num{8} hours. 

\section{System Model}
\label{section:system_model}

There is a total of $K$ single-antenna devices and a total of $M$ co-located \gls{bs} antennas. The set of active users trying to access the network is denoted by $\mathcal{K}_a$, with $|\mathcal{K}_a|\leq K$. To access the network, each active device $k\in\mathcal{K}_a$ sends a unique, non-orthogonal preamble of length $T$, known to the network. The pilot symbol of the preamble sent by device $k$ at pilot symbol $t$ is denoted by $s_{k,t}$. The channel vector between the receive antenna $m$ and user $k$ is denoted by $h_{k,m}\in \mathbb{C}$, and is considered fixed over the preamble duration. The received symbol at the $m$-th \gls{bs} antenna at time $t$ is
\begin{align}\label{eq:received_symbol}
    {y}_{t,m}=\sum_{k=0}^{K-1}{h}_{k,m} s_{k,t} \gamma_k + {w}_{t,m}, %\quad m=0,...,M-1,\ t=0,...,T-1,
\end{align}
where $\gamma_k$ is an unknown complex scalar and ${w}_{t,m}\in \mathbb{C}$ is \gls{iid} \gls{zmcscg} noise with variance $\sigma^2$. The unknown complex scalar $\gamma_k$ can be developed as
\begin{align}\label{eq:definition_gamma_k}
    \gamma_k= \sqrt{\rho_k} a_k e^{j\phi_k}, 
\end{align} 
where $\rho_k \in \mathbb{R}^+$ is the transmit power of device $k$, $a_k \in \left\{0,1\right\}$ is the device activity and $\phi_k$ models a potential phase offset.  This offset $\phi_k$ can account for a \gls{cfo}, where the phase is considered constant over the preamble duration. \update{This occurs when no frequency drift has occurred during the preamble duration, which is a feasible assumption as the \gls{cfo} is typically low, yielding negligible phase rotations during the preamble interval.} By assuming that all \(M\) antennas are perfectly synchronized, this offset is only dependent on the device. In case the device is inactive, $\gamma_k$ will be zero. We will introduce the term \textit{activity indicator} to denote $\gamma_k$.

Let us consider that the \gls{bs} knows a part of the \gls{csi}, \textit{i.e.}, ${g}_{k,m}$ in
\begin{align}\label{eq:prior_CSI_channel_model}
{h}_{k,m}={g}_{k,m}+ \lambda_k \epsilon_{k,m}, %\quad m=0,...,M-1,\ k=0,...,K-1, 
\end{align}
where $\epsilon_{k,m}$ are \gls{iid} \gls{zmcscg} variables with unit variance, and $\lambda_k \in \mathbb{R}^+$ models the unknown part of the \gls{csi}. The large-scale fading coefficient of user $k$ is $\beta_k  = \mathbb{E}(\norm{\vect{h}_k}^2/M) =\norm{\vect{g}_k}^2/M+\lambda_k^2$. The factor $\lambda_k$ models the quality of the known \gls{csi}. Hence, it quantifies the correlation of the actual channel ${h}_{k,m}$ to the known partial \gls{csi} ${g}_{k,m}$. In the extreme case with $\lambda_k=0$, the \gls{csi} is perfectly known and there is no uncertainty left, as was studied in~\cite{callebaut2021grant}. This could be the case in a fully static environment and if the \gls{csi} estimates are noiseless. However, for a realistic \gls{iot} scenario, even for static devices, \gls{csi} is not perfect due to i) environment dynamics and ii) noisy estimates. The parameter $\lambda_k$ then quantifies this imperfection. It is here assumed to be known\footnote{It could be set to a certain value depending on the user activity profile and/or tracked for each user over time.}. Another extreme case, as considered in \cite{GanesanTCOM,FenglerTIT}, is obtained when ${g}_{k,m}={0}, \forall m$, implying that only the large scale fading coefficient of user $k$ is known, \textit{i.e.}, $\lambda_k=\sqrt{\beta_k}$.

\section{Device Activity Detection}%
\label{section:Device_Activity_Detection}

This section describes the proposed activity detection algorithm. First, the log-likelihood of the received preamble is derived. Then, given its non-convex expression, an iterative approach is proposed to estimate the parameters $\gamma_k\ \forall k$. Finally, activity detection is performed.

\subsection{Log-Likelihood of the Received Symbols}

Combining (\ref{eq:received_symbol}) and (\ref{eq:prior_CSI_channel_model}), the symbol, received at \gls{bs} antenna $m$ and for pilot symbol $t$, is given by
\begin{align*}
	y_{t,m}&=\sum_{k=0}^{K-1} ({g}_{k,m} + {\epsilon}_{k,m} \lambda_k  )  s_{t,k} \gamma_k + {w}_{t,m}.
\end{align*}
Stacking the observations at antenna $m$ gives
\begin{align*}
\vect{y}_{m}&=\sum_{k=0}^{K-1} {g}_{k,m} \vect{s}_{k} \gamma_k + \sum_{k=0}^{K-1} {\epsilon}_{k,m} \lambda_k  \vect{s}_{k} \gamma_k  + \vect{w}_{m},
\end{align*}
where
\begin{align*}
    \vect{y}_m&=\begin{pmatrix} y_{0,m}\\ \vdots \\ y_{T-1,m} \end{pmatrix},\ \vect{s}_k=\begin{pmatrix} s_{0,k}\\ \vdots \\ s_{T-1,k} \end{pmatrix},\ \vect{w}_m=\begin{pmatrix} w_{m,0}\\ \vdots \\ w_{T-1,m} \end{pmatrix}.
\end{align*}
For a given value of $\gamma_k$, $\vect{y}_{m}|\gamma_k$ has a circularly symmetric Gaussian distribution with mean $\sum_{k=0}^{K-1} {g}_{k,m} \vect{s}_{k} \gamma_k $. After defining the vector $
\vect{\theta}_m =\sum_{k=0}^{K-1} {\epsilon}_{k,m} \lambda_k  \vect{s}_{k} \gamma_k  + \vect{w}_{m}$, the covariance matrix is
\begin{align}\label{eq:covariance_matrix_C}
\mat{C}&=\mathbb{E}\left( \vect{\theta}_m \vect{\theta}^H_m\right) =\sum_{k=0}^{K-1}\lambda_k^2|\gamma_k|^2 \vect{s}_k\vect{s}_k^H +\sigma^2\mat{I}_T, 
\end{align}
where we used the fact that ${\epsilon}_{k,m}$ were assumed to be \gls{iid} and the additive noise is white. Note that this covariance matrix does not depend on the antenna index $m$ and is thus valid for all $\vect{y}_m$.
% \begin{align*}
% \mat{C}&= \mathbb{E}(| \vect{y}_{m}- \mathbb{E}(\vect{y}_{m})|^2)\\
% &= \mathbb{E}\left(\left|\sum_{k=0}^{K-1} {g}_{k,m} \vect{s}_{k} \gamma_k + {\epsilon}_{k,m} \lambda_k  \vect{s}_{k} \gamma_k  + \vect{w}_{m} - \sum_{k=0}^{K-1} {g}_{k,m} \vect{s}_{k} \gamma_k\right|^2\right )\\
% &= \mathbb{E}\left(\left|{\epsilon}_{k,m} \lambda_k  \vect{s}_{k} \gamma_k  + \vect{w}_{m}\right|^2 \right)\\
% &= \mathbb{E}\left(\left({\epsilon}_{k,m} \lambda_k  \vect{s}_{k} \gamma_k  + \vect{w}_{m}\right)
% \left({\epsilon}_{k,m} \lambda_k  \vect{s}_{k} \gamma_k  + \vect{w}_{m}\right)^H
% \right)\\
% \end{align*}
Defining the vector $\vect{\gamma}=(\gamma_0,...,\gamma_{K-1})^T$, and $\vect{\Theta}_m=\vect{y}_m - \sum^{K-1}_{k=0} g_{k,m} \vect{s}_{k} \gamma_k$ the log-likelihood of the observation vector $\vect{y}_m$ is
\begin{align*}
\log p(\vect{y}_{m}|\vect{\gamma})&=-\ln\left(\left|\mat{C}\right|\right)-T\ln(\pi)
-\vect{\Theta}_m^H
\mat{C}^{-1} \vect{\Theta}_m.
\end{align*}
Given the conditional independence of ${\epsilon}_{k,m}$ and ${w}_{t,m}$ over the antennas, the different $\vect{y}_m$ are independent as well. Hence, the log-likelihood of the aggregated observations at all antennas $\vect{y}=(\vect{y}_0^T,...,\vect{y}_{M-1}^T)^T$ becomes 
\begin{align}\label{eq:log_likelihood}
&\log p(\vect{y}|\vect{\gamma})=
%\sum_{m=0}^{M-1} \log p(\vect{y}_m|\vect{\gamma})\label{eq:log_likelihood}\\
%&=
-M\ln\left(\left|\mat{C}\right|\right)-MT\ln(\pi)-\sum_{m=0}^{M-1}\vect{\Theta}_m^H
\mat{C}^{-1}
\vect{\Theta}_m.%\\
% &=-M\ln\left(\left|\mat{C}\right|\right)-MT\ln(\pi)-\sum_{m=0}^{M-1}\left(\vect{y}_{m}-\mat{\Gamma}_m \vect{\gamma}\right)^H
% \mat{C}^{-1}
% \left(\vect{y}_{m}-\mat{\Gamma}_m \vect{\gamma}\right),
\end{align}
% where we defined
% \begin{align*}
% 	\mat{\Gamma}_m&=\begin{pmatrix}
% 	\vect{s}_0 & ... & \vect{s}_{K-1}
% 	\end{pmatrix}\diag(g_{m,0},...,g_{k,m-1}).
% %	\mat{Y}&=\begin{pmatrix}
% %	\vect{y}_0 & ... & \vect{y}_{M-1}
% %	\end{pmatrix}
% \end{align*}

% One of the difference with the approaches in \cite{GanesanTCOM,FenglerTIT} is that we consider complex coefficients $\gamma_k$ and not simply real. This changes the optimization procedure and I am not sure we can use their simple trick... Another main difference is that we do not have just $\vect{y}_m$ but $\vect{y}_m$ minus something that depends on $\gamma_k$.

\begin{figure*}[t!]
\begin{align}
f(r_{k'})
&=-M\ln\left(\left|\mat{C}\right|\right)-\sum_{m=0}^{M-1}\vect{y}_{k',m}^H
\mat{C}^{-1}
\vect{y}_{k',m} -r_{k'}^2
\vect{s}_{k'}^H\mat{C}^{-1}\vect{s}_{k'} \sum_{m=0}^{M-1}|g_{k',m}|^2 +2\left|\sum_{m=0}^{M-1}\vect{y}_{k',m}^H
\mat{C}^{-1} \vect{s}_{k'} {g}_{k',m}\right|r_{k'}\label{eq:f_r_k_prime}
\end{align}%\\
\begin{align}
f(r_{k'},\phi_{k'})&=-M\ln\left(\left|\mat{C}\right|\right)-\sum_{m=0}^{M-1}\left(\vect{y}_{k',m}-{g}_{k',m} \vect{s}_{k'} r_{k'}e^{\jmath\phi_{k'}}\right)^H \mat{C}^{-1}
\left(\vect{y}_{k',m}-{g}_{k',m} \vect{s}_{k'} r_{k'}e^{\jmath\phi_{k'}}\right)\label{eq:f_r_k_phi}
\end{align}
\noindent\makebox[\linewidth]{\rule{\linewidth}{0.3pt}}
%\bottomrule%
\end{figure*}

\subsection{Iterative Algorithm for Maximizing Likelihood}

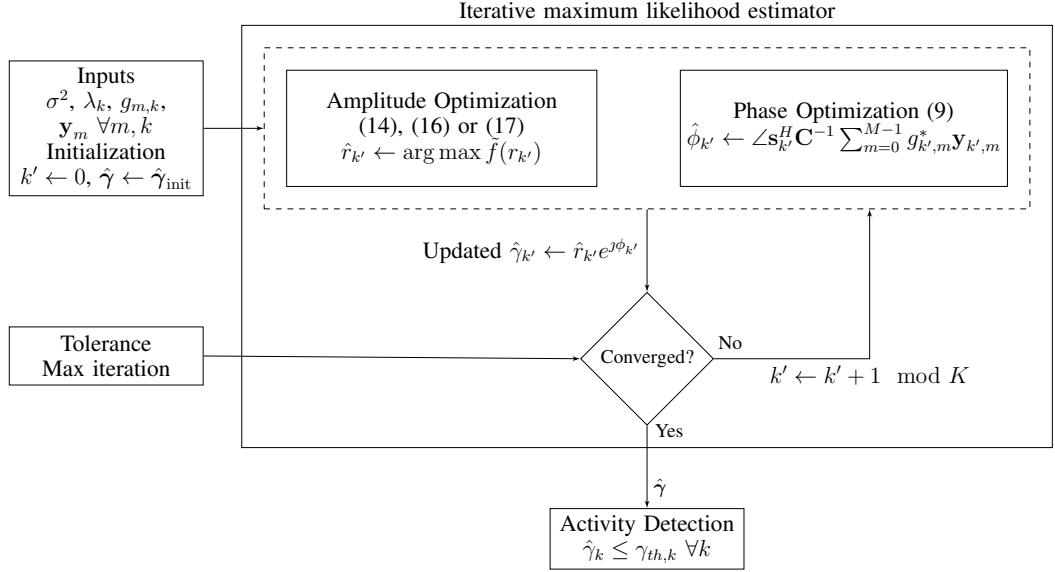
\begin{figure*}[t!]
	\centering 
\resizebox{0.8\linewidth}{!}{%

\tikzstyle{block} = [draw, rectangle, 
    minimum height=3em, minimum width=6em, minimum height = 3em, minimum width = 10em]
\tikzstyle{sum} = [draw, fill=blue!20, circle, node distance=1cm]
\tikzstyle{input} = [coordinate]
\tikzstyle{output} = [coordinate]
\tikzstyle{pinstyle} = [pin edge={to-,thin,black}]
\tikzstyle{container} = [draw, rectangle, inner sep=0.4cm]
\tikzstyle{decision} = [diamond, draw, text badly centered, inner sep=3pt]

\begin{tikzpicture}[auto, node distance=1.5cm,>=latex']

    % We start by placing the blocks
    \node [block, name=input, align=center, font=\large] (input) {\large Inputs\\ $\sigma^2$, $\lambda_k$, $g_{m,k}$,\\ $\vect{y}_m$ $\forall m,k$ \\ 
    Initialization \\ $k' \gets 0$, $\hat{\vect{\gamma}} \gets \hat{\vect{\gamma}}_{\mathrm{init}}$};

    \node [block, minimum height = 6em, minimum width=16em, align=center, right= of input, font=\large] (amplitude) {\large Amplitude   Optimization\\ (\ref{eq:update_partial_CSI}), (\ref{eq:update_no_CSI}) or (\ref{eq:update_prior_CSI}) \\ $\hat{r}_{k'} \gets \arg \max \tilde{f}(r_{k'})$};
    
    \node [block, align=center, minimum height = 6em, minimum width=16em, right= of amplitude, font=\large] (phase) {\large Phase  Optimization~(\ref{eq:phase_update})\\  $\hat{\phi}_{k'} \gets \angle  \vect{s}_{k'}^H \mat{C}^{-1} \sum_{m=0}^{M-1} {g}_{k',m}^*  \vect{y}_{k',m}$};

    %\path (amplitude) -- (phase) coordinate[midway] (aux);
    
    %\node [block, align=center, below of=aux] (update) {Update\\ $\hat{\gamma}_k = \hat{r}_{k'} e^{\jmath\phi_{k'}} $};

   \node [container,fit=(amplitude) (phase), label=below:{}, dashed] (inner_container) {};

   \node [decision, below=of inner_container] (converged) {Converged?};
   
   %\node [block, align=center, left= of converged] (iter) {$k' \gets k'+1 \mod K$};
    
   \node [container,fit=(converged) (inner_container), label=above:{\large Iterative maximum likelihood estimator}] (outer_container) {};

   \node [block, name=tolerance, align=center, left=of converged.center, below=2.9cm of input, font=\large, anchor=center] (tolerance) {\large Tolerance \\ Max iteration};

    \node [block, align=center, below=of converged, font=\large] (detection) {\large Activity Detection\\ $\hat{\gamma}_k \leq  \gamma_{th,k}\ \forall k$};

    \draw[->] (input.east) -- (inner_container.west);
    %\draw[->] (init.east) |- (inner_container.west);
    
    \draw[->] (tolerance.east) -- (converged.west);
    
    %\draw[->] (converged.west) -- node[midway]{No} (iter.east);

    \node [] at ([shift={(0.3,0.3)}]converged.east) {No};
    
    \node [] at ([shift={(0.4,-0.1)}]converged.south) {Yes};

    \draw[->] (converged.east) -|node[below, font=\large]{$k' \gets k'+1 \mod K$} (inner_container.-20);
    
     %\draw[->] (iter.north) -- (amplitude.south);
     
     \draw[->] (inner_container.south) --node[left, font=\large]{Updated $\hat{\gamma}_{k'} \gets \hat{r}_{k'} e^{\jmath\phi_{k'}} $} (converged.north);
    
    \draw[->] (converged.south) --node[near end]{$\hat{\vect{\gamma}}$} (detection.north);
    
\end{tikzpicture}%
}
	\caption{Block diagram of the iterative maximum likelihood estimator and activity detection.}\label{fig:iterative_algorithm} 
	%\vspace{-1em}
\end{figure*}

The maximum likelihood estimator of $\vect{\gamma}$ is obtained by maximizing
\begin{align*}
    \hat{\vect{\gamma}}_{\mathrm{ML}} &= \arg \max_{\vect{\gamma}} \log p(\vect{y}|\vect{\gamma}).
\end{align*}
This problem is not trivial to solve given the nonlinear and non-convex dependence of the log-likelihood, more specifically the covariance matrix $\mat{C}$, in $\vect{\gamma}$. 
An idea to maximize the likelihood is to use an iterative approach, similarly as \cite{GanesanTCOM,FenglerTIT}: at each iteration, all $\gamma_k$ are kept fixed but one, which is optimized and updated. This way, they get updated one by one until convergence is attained, \textit{i.e.}, a maximum number of iterations or a certain tolerance is reached. A block diagram of the algorithm is given in Figure~\ref{fig:iterative_algorithm} and the pseudocode is summarized in Algorithm~\ref{alg:cap}.

% \begin{figure}[t!]
% 	\centering 
% 	\resizebox{0.8\linewidth}{!}{%
% 		{\includegraphics[clip, trim=0cm 8.5cm 12.5cm 0cm, scale=1]{Fig/iterative_algorithm.pdf}} %gauche bas droite haut
% 	}
	
% 	\caption{Block diagram of the iterative maximum likelihood estimator and activity detection.\TODO{add cross refs to equation numbers for the optimizations}}\label{fig:iterative_algorithm} 
% 	\vspace{-1em}
% \end{figure}

Let us consider that the complex-valued $\gamma_{k'}$ needs to be updated. Using the definition introduced in (\ref{eq:definition_gamma_k}), we can rewrite $\gamma_{k'}$ with a phase-amplitude decomposition: $\gamma_{k'}=r_{k'}e^{\jmath\phi_{k'}}$, with $r_{k'}=|\gamma_{k'}|=\sqrt{\rho_{k'}}a_{k'}$. The optimization with respect to $\gamma_{k'}$ is done in the following in several steps: i)~optimizing the phase $\phi_{k'}$ for a fixed value of $r_{k'}$, ii)~re-inserting this expression in the objective function to remove the dependence in~$\phi_{k'}$ and iii)~optimizing the amplitude $r_{k'}$. 

\subsubsection{Phase optimization}

To highlight the dependence of the objective function in $\gamma_{k'}$ for constant values of other~$\gamma_{k},\ k\neq k'$, let us define the vector
\begin{align}
	\vect{y}_{k',m}&=\vect{y}_{m}-\sum_{k=0, k\neq k'}^{K-1} {g}_{k,m} \vect{s}_{k} \gamma_k, \label{eq:y_m_k_prime}
\end{align}
which can be seen as a cancellation of device interference to isolate the contribution from device~$k'$. Hence, the objective function to maximize can be written as in (\ref{eq:f_r_k_phi})\footnote{For clarity, we omit in the following the constant term $MT\ln(\pi)$ which does not affect optimization as it does not depend on $\vect{\gamma}$ and vanishes after differentiation.}.

where we explicitly express the dependence in $(r_{k'},\phi_{k'})$ while the other $(r_{k},\phi_{k}),k\neq k'$ do not appear since they are considered constant. Note that the matrix $\mat{C}$, defined in (\ref{eq:covariance_matrix_C}), does not depend on $\phi_{k'}$ but only $r_{k'}$. In the extreme case of no prior \gls{csi}, \textit{i.e.}, $g_{k',m}=0$ $\forall m$, the dependence of $f(r_{k'},\phi_{k'})$ in $\phi_{k'}$ disappears and there is an underdetermination and no estimate of the phase offset can be obtained. In other cases, we can find that, after some manipulations,
\begin{align}
	%\frac{df}{d\phi_{k'}}=0\leftrightarrow
 \hat{\phi}_{k'}&=\angle  \vect{s}_{k'}^H \mat{C}^{-1}      \sum_{m=0}^{M-1}{g}_{k',m}^*  \vect{y}_{k',m}. \label{eq:phase_update}
\end{align}
This result has an intuitive understanding: the optimal phase $\phi_k'$ tends to align the partial \gls{csi} with the observations due to device $k'$. 

\subsubsection{Removing the phase dependence}

Inserting this optimal value in the objective function $f(r_{k'},\phi_{k'})$ makes the dependence in $\phi_{k'}$ vanish and gives (\ref{eq:f_r_k_prime}).

\subsubsection{Amplitude optimization}

To alleviate the dependence on $r_{k'}$ in (\ref{eq:f_r_k_prime}), 
% In the expression of $f(r_{k'})$, a complex modulus of an expression that depends on $r_{k'}$ has appeared, which complicates differentiation. To alleviate this and make the dependence in $r_{k'}$ more clear, 
let us define
\begin{align}
	\mat{C}_{-k'}&=\mat{C}-\lambda_{k'}^2|\gamma_{k'}|^2 \vect{s}_{k'}\vect{s}_{k'}^H\label{eq:C_minus_k_prime}\\
	&=\sum_{k\setminus k'}\lambda_k^2|\gamma_k|^2 \vect{s}_k\vect{s}_k^H +\sigma^2\mat{I}_T, \nonumber
\end{align}
which does not depend on $r_{k'}$ and is full rank, thus invertible. Applying the Sherman-Morrison formulas~\cite{sherman1950adjustment}
% \begin{align}
% 	\left(\mat{A} + \vect{u}\vect{v}^\textsf{T}\right)^{-1} &= \mat{A}^{-1} - \frac{\mat{A}^{-1}\vect{u}\vect{v}^\textsf{T}\mat{A}^{-1}}{1 + \vect{v}^\textsf{T}\mat{A}^{-1}\vect{u}}\nonumber\\
% 	\left(\mat{A} + \vect{u}\vect{v}^\textsf{T}\right)^{-1}\vect{u} %&= \mat{A}^{-1}\vect{u} - \frac{\mat{A}^{-1}\vect{u}\vect{v}^\textsf{T}\mat{A}^{-1}\vect{u}}{1 + \vect{v}^\textsf{T}\mat{A}^{-1}\vect{u}}
% 	&= \frac{\mat{A}^{-1}\vect{u}}{1 + \vect{v}^\textsf{T}\mat{A}^{-1}\vect{u}} \label{eq:Sherman_Morrison_formula}
% \end{align}	
to $\mat{C}^{-1}$ and $\mat{C}^{-1}\vect{s}_{k'}$ gives
\begin{align}
	\mat{C}^{-1}& =\mat{C}_{-k'}^{-1}-\frac{\mat{C}_{-k'}^{-1}\vect{s}_{k'}\vect{s}_{k'}^H\mat{C}_{-k'}^{-1}r_{k'}^2\lambda_{k'}^2}{1+\vect{s}_{k'}^H\mat{C}_{-k'}^{-1}\vect{s}_{k'}r_{k'}^2\lambda_{k'}^2}\label{eq:C_inverse_Sherman_Morrison_formula}%\\
% 	\mat{C}^{-1}\vect{s}_{k'}r_{k'}\lambda_{k'}&=\frac{\mat{C}_{-k'}^{-1}\vect{s}_{k'}r_{k'}\lambda_{k'}}{1+\vect{s}_{k'}^H\mat{C}_{-k'}^{-1}\vect{s}_{k'}r_{k'}^2\lambda_{k'}^2}\\
	%\mat{C}^{-1}\vect{s}_{k'}&=\frac{\mat{C}_{-k'}^{-1}\vect{s}_{k'}}{1+\vect{s}_{k'}^H\mat{C}_{-k'}^{-1}\vect{s}_{k'}r_{k'}^2\lambda_{k'}^2} \label{eq:C_inverse_s_Sherman_Morrison_formula}
\end{align}
% Inserting these expression in (\ref{eq:f_r_k_prime}), the objective function becomes
% \begin{align*}
% f(r_{k'})%&=-M\ln\left(\left|\mat{C}\right|\right)-MT\ln(\pi)\\
% % &-\sum_{m=0}^{M-1}\vect{y}_{k',m}^H
% % \mat{C}^{-1}
% % \vect{y}_{k',m}-r_{k'}^2
% % \vect{s}_{k'}^H\mat{C}^{-1}\vect{s}_{k'} \sum_{m=0}^{M-1}|g_{k',m}|^2+2\left|  \frac{\sum_{m=0}^{M-1}\vect{y}_{k',m}^H\mat{C}_{-k'}^{-1}\vect{s}_{k'}{g}_{k',m}}{1+\vect{s}_{k'}^H\mat{C}_{-k'}^{-1}\vect{s}_{k'}r_{k'}^2\lambda_{k'}^2} \right|r_{k'}\\
% &=-M\ln\left(\left|\mat{C}\right|\right)-\sum_{m=0}^{M-1}\vect{y}_{k',m}^H\mat{C}^{-1}\vect{y}_{k',m}\\
% &-r_{k'}^2 \vect{s}_{k'}^H\mat{C}^{-1}\vect{s}_{k'} \sum_{m=0}^{M-1}|g_{k',m}|^2+2  \frac{\left|\sum_{m=0}^{M-1}\vect{y}_{k',m}^H\mat{C}_{-k'}^{-1}\vect{s}_{k'}{g}_{k',m}\right|}{1+\vect{s}_{k'}^H\mat{C}_{-k'}^{-1}\vect{s}_{k'}r_{k'}^2\lambda_{k'}^2}r_{k'}
% \end{align*}
% Good news, the differenciability problem seems to be solved. Now we apply the Sherman-Morrison formula on each inverse to put in evidence the dependence in $r_{k'}$. 
We insert these expressions in (\ref{eq:f_r_k_prime}) and we omit terms that do not depend on $r_{k'}$, which will vanish after taking the derivative. This gives
\begin{align}
\tilde{f}(r_{k'})%&=-M\ln\left(\left|\mat{C}\right|\right)+\sum_{m=0}^{M-1}\vect{y}_{k',m}^H
% \frac{\mat{C}_{-k'}^{-1}\vect{s}_{k'}\vect{s}_{k'}^H\mat{C}_{-k'}^{-1}r_{k'}^2\lambda_{k'}^2}{1+\vect{s}_{k'}^H\mat{C}_{-k'}^{-1}\vect{s}_{k'}r_{k'}^2\lambda_{k'}^2}
% \vect{y}_{k',m}\\
% &-r_{k'}^2
% \frac{\vect{s}_{k'}^H\mat{C}_{-k'}^{-1}\vect{s}_{k'}}{1+\vect{s}_{k'}^H\mat{C}_{-k'}^{-1}\vect{s}_{k'}r_{k'}^2\lambda_{k'}^2} \sum_{m=0}^{M-1}|g_{k',m}|^2+2  \frac{\left|\sum_{m=0}^{M-1}\vect{y}_{k',m}^H\mat{C}_{-k'}^{-1}\vect{s}_{k'}{g}_{k',m}\right|}{1+\vect{s}_{k'}^H\mat{C}_{-k'}^{-1}\vect{s}_{k'}r_{k'}^2\lambda_{k'}^2}r_{k'}\\
%&=-M\ln\left(\left|\mat{C}\right|\right)+\sum_{m=0}^{M-1}
% \frac{|\vect{y}_{k',m}^H\mat{C}_{-k'}^{-1}\vect{s}_{k'}|^2\lambda_{k'}^2r_{k'}^2}{1+\vect{s}_{k'}^H\mat{C}_{-k'}^{-1}\vect{s}_{k'}r_{k'}^2\lambda_{k'}^2}\\
%&-r_{k'}^2\frac{\vect{s}_{k'}^H\mat{C}_{-k'}^{-1}\vect{s}_{k'}}{1+\vect{s}_{k'}^H\mat{C}_{-k'}^{-1}\vect{s}_{k'}r_{k'}^2\lambda_{k'}^2} \sum_{m=0}^{M-1}|g_{k',m}|^2+2  \frac{\left|\sum_{m=0}^{M-1}\vect{y}_{k',m}^H\mat{C}_{-k'}^{-1}\vect{s}_{k'}{g}_{k',m}\right|}{1+\vect{s}_{k'}^H\mat{C}_{-k'}^{-1}\vect{s}_{k'}r_{k'}^2\lambda_{k'}^2}r_{k'}\\
&=-M\ln\left(\left|\mat{C}\right|\right)+\frac{\alpha r_{k'}^2+\beta r_{k'}}{1+\delta r_{k'}^2}, \label{eq:f_tilde}
\end{align}
where we defined the constants (independent of $r_{k'}$) $\alpha$, $\beta$ and $\delta$, as
\begin{align}
	\alpha&=\sum_{m=0}^{M-1}
	{|\vect{y}_{k',m}^H\mat{C}_{-k'}^{-1}\vect{s}_{k'}|^2\lambda_{k'}^2}-
	{\vect{s}_{k'}^H\mat{C}_{-k'}^{-1}\vect{s}_{k'}} \sum_{m=0}^{M-1}|g_{k',m}|^2\nonumber\\
	\beta&=2  \left|\sum_{m=0}^{M-1}\vect{y}_{k',m}^H\mat{C}_{-k'}^{-1}\vect{s}_{k'}{g}_{k',m}\right|\nonumber\\
	\delta&=\vect{s}_{k'}^H\mat{C}_{-k'}^{-1}\vect{s}_{k'}\lambda_{k'}^2. \label{eq:alpha_beta_gamma}
\end{align}
One can note that $\tilde{f}(r_{k'})$ in (\ref{eq:f_tilde}) can now be differentiated with respect to $r_{k'}$, using the differential rule $\partial (\log |\mat{A}|)=\tr[\mat{A}^{-1} \partial \mat{A}]$. Setting the derivative to zero gives, noting that the denominator is always strictly positive,
\begin{align}
	% \frac{d\tilde{f}}{dr_{k'}}&=0\nonumber\\
	%\leftrightarrow
 0&=-r_{k'}^32M\delta^2-r_{k'}^2\beta\delta+r_{k'}(-2M\delta+2\alpha)+\beta, \label{eq:update_partial_CSI}
\end{align}
which is a polynomial of degree~3 in $r_{k'}$. There are closed-form solutions for the roots of such polynomials. Following \textit{Descarte's rule of signs}, (\ref{eq:update_partial_CSI}) has only one real and positive root, as required, in case the terms are non-zero. Below we discuss the special cases when the terms are not non-zero, i.e., when there is no or complete \gls{csi} knowledge.

%One should still check for the best solutions among them, restricted to be positive and real.\footnote{even though we could not prove it analytically, in all our simulations only one such root was found, solving this root selection problem.}

The algorithm is summarized in the pseudocode \textbf{Algorithm~\ref{alg:cap}}. At each iteration, the constants $\alpha$, $\beta$ and $\delta$ can be easily re-evaluated based on~(\ref{eq:alpha_beta_gamma}). They require the matrix inversion $\mat{C}_{-k'}^{-1}$. To avoid re-computing a full inverse at each iteration, one can rely on the Sherman-Morrison formula and on the current knowledge of $\mat{C}^{-1}$, which is updated at the end of each iteration by inserting the obtained value of $r_{k'}$ in~(\ref{eq:C_inverse_Sherman_Morrison_formula}). Using~(\ref{eq:C_minus_k_prime}), we find
% \begin{align}
%     \mat{C}_{-k'}^{-1}\vect{s}_{k'}&=\left(\mat{C}-\lambda_{k'}^2|\gamma_{k'}|^2 \vect{s}_{k'}\vect{s}_{k'}^H\right)^{-1}\vect{s}_{k'} \nonumber\\
% 	&= \frac{\mat{C}^{-1}\vect{s}_{k'}}{1 -  \lambda_{k'}^2|\gamma_{k'}|^2 \vect{s}_{k'}^H \mat{C}^{-1}\vect{s}_{k'}}. \label{eq:C_minus_k_prime_Sherman_Morrison}
% \end{align}
\begin{align}
    \mat{C}_{-k'}^{-1}
    %&=\left(\mat{C}-\lambda_{k'}^2|\gamma_{k'}|^2 \vect{s}_{k'}\vect{s}_{k'}^H\right)^{-1}\nonumber\\
	&= \mat{C}^{-1}+\frac{\lambda_{k'}^2|\gamma_{k'}|^2 \mat{C}^{-1}\vect{s}_{k'}\vect{s}_{k'}^H\mat{C}^{-1} }{1 -  \lambda_{k'}^2|\gamma_{k'}|^2 \vect{s}_{k'}^H \mat{C}^{-1}\vect{s}_{k'}}. \label{eq:C_minus_k_prime_Sherman_Morrison}
\end{align}
Moreover, computations can be optimized as several quantities appear multiple times and can be computed only once. \update{The complexity of the proposed algorithm is $\mathcal{O}(IMT^2)$, where $I$ is the number of iterations.\footnote{Note that $I$ will in practice depend on $K$ as we will iterate $N$ times over all users $K$, but this is not a requirement.}}

We now investigate two particular cases, to gain further insights.

\begin{algorithm}
\caption{Iterative maximum likelihood device activity detector}\label{alg:cap}
\small 
\begin{algorithmic}
\Require $\sigma^2,\lambda_k,\vect{y}_m, g_{k,m}, \hat{\vect{\gamma}}_{\mathrm{init}} \ \forall k,m$
\State $k' \gets 0$
\State $\hat{\vect{\gamma}} \gets \hat{\vect{\gamma}}_{\mathrm{init}}$
\State $\mat{C}^{-1} \gets \left(\sum_{k=0}^{K-1}\lambda_k^2|\hat{\gamma}_k|^2 \vect{s}_k\vect{s}_k^H +\sigma^2\mat{I}_T\right)^{-1}$
\While{Not converged}

    Compute $\vect{y}_{k',m}$, $\mat{C}_{-k'}^{-1}$, $\alpha$, $\beta$ and $\delta$ based on (\ref{eq:y_m_k_prime}), (\ref{eq:C_minus_k_prime_Sherman_Morrison}) and (\ref{eq:alpha_beta_gamma})

    \State $\hat{r}_{k'} \gets \arg \max \tilde{f}(r_{k'})$ \Comment{Update amplitude based on (\ref{eq:update_partial_CSI}), (\ref{eq:update_no_CSI}) or (\ref{eq:update_prior_CSI})}
    \State $\hat{\phi}_{k'} \gets \angle  \vect{s}_{k'}^H \mat{C}^{-1}      \sum_{m=0}^{M-1} {g}_{k',m}^*  \vect{y}_{k',m}$ \Comment{Update phase}
    \State $\hat{\gamma}_{k'} \gets \hat{r}_{k'}e^{\jmath \hat{\phi}_{k'}}$
    \State $\mat{C}^{-1} \gets \mat{C}_{-k'}^{-1}-\frac{\mat{C}_{-k'}^{-1}\vect{s}_{k'}\vect{s}_{k'}^H\mat{C}_{-k'}^{-1}r_{k'}^2\lambda_{k'}^2}{1+\vect{s}_{k'}^H\mat{C}_{-k'}^{-1}\vect{s}_{k'}r_{k'}^2\lambda_{k'}^2}$
    \State $k' \gets k'+1 \mod K$
\EndWhile
\end{algorithmic}
\end{algorithm}

\paragraph{Particular case: device with no \gls{csi}}

Now consider that, for a given $k'$, ${g}_{k',m}={0}\  \forall m$. This could be because this device is new or moving a lot, such that its \gls{csi} is outdated. Only, its parameter $\lambda_{k'}$ is known, which is equal to the large-scale fading coefficient $\sqrt{\beta_{k'}}$. At iteration of user $k'$, evaluating (\ref{eq:alpha_beta_gamma}) for ${g}_{k',m}={0}\  \forall m$ implies that $\beta=0$. Hence, (\ref{eq:update_partial_CSI}) simplifies to
\begin{align*}
	0=2r_{k'}(-r_{k'}^2M\delta^2-M\delta+\alpha),
\end{align*}
which has a trivial solution in $r_{k'}=0$. One of the other roots is always negative. Keeping only the positive one, we find the amplitude update
\begin{align}
\hat{r}_{k'}=\sqrt{\frac{\alpha-M\delta}{M\delta^2}}. \label{eq:update_no_CSI}
\end{align}
If this root is imaginary, we set $\hat{r}_{k'}$ to zero. As discussed before introducing the phase update equation~(\ref{eq:phase_update}), in the case of no prior \gls{csi}, the phase ambiguity cannot be resolved. This particular case gives an update relatively similar to the maximum likelihood estimator derived in~\cite[(23)]{FenglerTIT}, where their \gls{ml} expression estimates the error, while ours estimates directly the coefficient $\hat{r}_{k'}$, given the same estimate of $\gamma_{k'}$ at each iteration.

\paragraph{Particular case: device with complete prior \gls{csi}}

Now, consider that, for a given $k'$, $\lambda_{k'}={0}$, so that the \gls{csi} is perfectly known. Only the phase shift and the transmit power are unknown. At iteration of user $k'$, evaluating (\ref{eq:alpha_beta_gamma}) for $\lambda_{k'}={0}$ implies that $\delta=0$. Hence, (\ref{eq:update_partial_CSI}) simplifies to a linear equation $0=r_{k'}2\alpha+\beta$, which gives the following amplitude update
\begin{align}
\hat{r}_{k'}&=\frac{-\beta}{2\alpha}=\frac{| \vect{s}_{k'}^H
	\mat{C}^{-1} 
	\sum_{m=0}^{M-1}{g}_{k',m}^*\vect{y}_{k',m}|}{\vect{s}_{k'}^H
	\mat{C}^{-1} 
	\vect{s}_{k'} \sum_{m'}|{g}_{k',m'}|^2   },\label{eq:update_prior_CSI}
\end{align}
while the phase is updated according to (\ref{eq:phase_update}).

% \TODO{this is similar to our conference paper, but for a single device. Not sure this case really happens in practice since there is always a bit or noise/mobility on channel estimate. Even if correlation is very high like 0.99, not perfect...}

% At iteration of user $k'$, the objective function to be maximized is
% \begin{align*}
% f(\gamma_{k'})&=-M\ln\left(\left|\mat{C}\right|\right)-\sum_{m=0}^{M-1}\left(\vect{y}_{k',m}-{g}_{k',m} \vect{s}_{k'} \gamma_{k'}\right)^H
% \mat{C}^{-1}
% \left(\vect{y}_{k',m}-{g}_{k',m} \vect{s}_{k'} \gamma_{k'}\right),
% \end{align*}
% where we kept the dependence on the complex variable $\gamma_{k'}$ on purpose. Indeed, one can check that this function is concave in $\gamma_{k'}$. Indeed, $\mat{C}$, defined in (\ref{eq:covariance_matrix_C}), does not depend on $\gamma_{k'}$ given that $\lambda_{k'}=0$. Hence, $f(\gamma_{k'})$ is a quadratic function in $\gamma_{k'}$. Taking the Wirtinger derivative of $f(\gamma_{k'})$ with respect to $\gamma_{k'}^*$ and setting it to zero gives
% \begin{align}
% \frac{df}{d\gamma_{k'}^*}&=0%\\
% % 0&=\sum_{m=0}^{M-1}{g}_{k',m}^* \vect{s}_{k'}^H
% % \mat{C}^{-1} 
% % \left(\vect{y}_{k',m}-{g}_{k',m} \vect{s}_{k'} \gamma_{k'}\right)\\
% \leftrightarrow \hat{\gamma}_{k'}=\frac{ \vect{s}_{k'}^H
% 	\mat{C}^{-1} 
% 	\sum_{m=0}^{M-1}{g}_{k',m}^*\vect{y}_{k',m}}{\vect{s}_{k'}^H
% 	\mat{C}^{-1} 
% 	\vect{s}_{k'} \sum_{m'}|{g}_{k',m'}|^2   }. \label{eq:update_prior_CSI}
% \end{align}

\subsubsection{Initialization}

To start the iterative algorithm, we consider different choices to initialize $\hat{\vect{\gamma}}_{\mathrm{init}}$. A simple choice is to initialize to zero, \textit{i.e.}, $\hat{\vect{\gamma}}_{\mathrm{init}}^{0}=\vect{0}$. Another choice is to initialize solely based on the available prior \gls{csi}, considering that $\lambda_k \approx 0,\ \forall k$. The estimator is similar to \cite{callebaut2021grant}, except that, here, prior \gls{csi} is used instead of complete \gls{csi}. 

If $\lambda_k \approx 0,\ \forall k$, the covariance matrix $\mat{C}$, defined in (\ref{eq:covariance_matrix_C}), simplifies to $\mat{C}=\sigma^{-2}\mat{I}_T$, which is independent of $\gamma_k$. Hence, many terms of the log-likelihood in (\ref{eq:log_likelihood}) become independent of $\vect{\gamma}$. Maximizing (\ref{eq:log_likelihood}) becomes equivalent to the following minimization
\begin{align*}
    \max_{\vect{\gamma}} \log p(\vect{y}|\vect{\gamma})&=\min_{\vect{\gamma}}\sum_{m=0}^{M-1}\left\|\vect{y}_{m}-\sum_{k=0}^{K-1} {g}_{k,m} \vect{s}_{k} \gamma_k\right\|^2\\
    &=\min_{\vect{\gamma}}\left\|\vect{y}-\mat{\Gamma} \vect{\gamma}\right\|^2,
\end{align*}
where we defined the vector and matrix notations
\begin{align*}
    &\vect{y}=\begin{pmatrix}
    \vect{y}_{0} \hdots  \vect{y}_{M-1}
    \end{pmatrix}^T,\ \mat{\Gamma}= \begin{pmatrix}
    \mat{\Gamma}_{0} \hdots   \vect{\Gamma}_{M-1}
    \end{pmatrix}^T,\\
    &\mat{\Gamma}_m=\begin{pmatrix}
    \vect{s}_{0} \hdots \vect{s}_{K-1}
    \end{pmatrix} \diag{ (g_{0,m} \hdots g_{{K-1},m})}.
\end{align*}
This minimization problem is a quadratic function of $\vect{\gamma}$, which is a least squares problem. The estimate has the following closed-form expression
\begin{align}
    \hat{\vect{\gamma}}_{\mathrm{init}}^{\mathrm{ZF}}&= \left(\mat{\Gamma}^H \mat{\Gamma}\right)^{-1} \mat{\Gamma}^H \vect{y}.
    \label{eq:zf}
\end{align}
This last estimator can be seen as a \gls{zf} estimator, which requires a matrix inversion. To avoid ill-conditioning, a first necessary condition is that $K\leq M T$. This condition is not sufficient as the channel and preamble of two devices could be correlated, especially when $K$ is on the order of $MT$. Moreover, if no prior information is available for a given user $k'$, \textit{i.e.}, $g_{k',m}=0,\ \forall m$, the inverse will also be ill-conditioned. This implies that the $k'$-th column of $\mat{\Gamma}$ becomes null and thus $\mat{\Gamma}$ is rank deficient. Moreover, the prior \gls{csi} might be noisy, leading to unstable results.

To make initialization more robust, we can use an \gls{lmmse} criterion. To do this, some prior knowledge must be assumed on the statistics of $\vect{\gamma}$, more specifically, its first and second order moments. We here make the following assumptions: i) the activity of each device is independent of one another, ii) the average activity and average transmit power of each device is known and iii) no prior information is known on the phase offset so that $\phi_k$ is considered uniformly distributed between 0 and $2\pi$. Under these assumptions, we have $\mathbb{E}(\vect{\gamma})=\vect{0}$ and $\mat{D}=\mathbb{E}(\vect{\gamma}\vect{\gamma}^H)=\diag\left(\mathbb{E}(a_0)\mathbb{E}(\rho_0),...,\mathbb{E}(a_{K-1})\mathbb{E}(\rho_{K-1})\right)$. Hence, for the linear observation model $\vect{y}=\mat{\Gamma} \vect{\gamma} + \vect{w}$, still considering that $\lambda_k \approx 0,\ \forall k$, the \gls{lmmse} estimator of $\vect{\gamma}$ is then given by \cite{kay1993fundamentals}
\begin{align}
\hat{\vect{\gamma}}_{\mathrm{init}}^{\mathrm{LMMSE}}&= \left(\mat{\Gamma}^H \mat{\Gamma} + \sigma^2 \mat{D}^{-1} \right)^{-1} \mat{\Gamma}^H  \vect{y}.
    \label{eq:lmmse}
\end{align}
Note that the matrix to be inverted is always well-conditioned. Finally, a \gls{mf} estimator could be used to avoid the need for matrix inversion.
\begin{align}
    \hat{\vect{\gamma}}_{\mathrm{init}}^{\mathrm{MF}}&= \left(\diag(\mat{\Gamma}^H \mat{\Gamma}) + \sigma^2 \mat{D}^{-1} \right)^{-1} \mat{\Gamma}^H  \vect{y}.
    \label{eq:mf}
\end{align}

\subsection{Activity Detection}

A non-negative activity threshold $\gamma_{\text{th},k}$ is applied for each device $k$. A device is considered active if $\abs{\hat{\gamma}_k} \geq \gamma_{th,k}$. 
The real-valued threshold is defined as,
\begin{equation}\label{eq:threshold}
    \gamma_{\text{th}, k} = v \sqrt{\text{SNR}_k}^{-1},
\end{equation}
where $v$ is chosen to have a desired probability of false alarm and miss detection performance and with $\text{SNR}_k = M\beta_k / \sigma^2  = \mathbb{E}(\norm{\vect{h}_k}^2) / \sigma^2  =(\norm{\vect{g}_k}^2+M\lambda_k^2) / \sigma^2$. 

Miss detection happens when a device was undetected, while it was actually transmitting. Equivalently, a false alarm occurs if a device is considered active by the algorithm but was not. We define the probability of miss detection as the average ratio of undetected devices to the number of active devices $
    \prob_{\mathrm{md}} = 1 - \left( \abs{\mathcal{K}_a \cap \hat{\mathcal{K}}_a } / \abs{\mathcal{K}_a} \right)$,
where $\mathcal{K}_a$ is the set of active devices and $\hat{\mathcal{K}}_a = \{k | \hat{a}_k=1, \forall \in [1,K]\}$ denotes the estimated set of active devices.
Note that on average $\abs{\mathcal{K}_a}=K_a$.
Similarly, the probability of false alarm is the ratio of inactive devices considered active to the number of inactive devices and is given by
    $\prob_{\mathrm{fa}} = 
            \left( \abs{\hat{\mathcal{K}}_a \setminus  \mathcal{K}_a}
            / (K - \abs{\mathcal{K}_a}) \right)$.

A trade-off can be made between the two probabilities by varying $v$ in~(\ref{eq:threshold}). A lower $v$ yields a lower activity threshold, resulting in more devices considered active. This in turn lowers the probability of miss detection, while increasing the probability of generating a false alarm. In the simulations, the parameter $v$ is swept across the range $[-40, 40]$\,\si{\deci\bel}. 

%To conveniently study the impact of different parameters, the metric \textit{error probability} is introduced $\prob_{e}=\prob_{\text{md}}=\prob_{\text{fa}}$.\footnote{\update{This metric is an approximation, where we numerically find $\prob_{\text{md}} \approx \prob_{\text{fa}}$ by finding the threshold where $\argmin{| \prob_{\text{md}} - \prob_{\text{fa}} | }$.}}

%\input{Section/Channel_Estimation_Decoding}

\section{Numerical Assessment}
\label{section:experimental_results}
\begin{table}[tbp]
    \centering
    \renewcommand*{\arraystretch}{1.2}
    \caption{Simulation parameter set with default values.}
    \label{tab:initial-access-setup}
    \centering
    \footnotesize
      \begin{threeparttable}
    \begin{tabular}{@{}lll@{}}
    \toprule
    Parameter &  Symbol & Default value\\ \midrule
    
    Number of devices & $K$ & \num{500}\\
    %Number of \glspl{ap} & $N$ & -\\
    Number of total \gls{bs} antennas & $M$ & 64\\
    \Acrlong{snr} &  \acrshort{snr}  & \SI{20}{\deci\bel}\\
    Device activity probability & $\epsilon_a$ & \num{0.1}\\
    % Transmit power of the device & $\rho$ & \SI{1}{\milli\watt}\\
    % Bandwidth & $B$ & \SI{125}{\kilo\hertz}\\
    % Thermal noise &$\sigma^2$ & \SI{-122.88}{\dBm} \\
    Pilot sequence & $s_k$ & $\cn{0}{1}$\\
    Pilot length &$\tau_p$  & \num{10} symbols\\
    Phase offset & $\phi_k$ &$\sim\mathcal{U}_{[0, 2\pi]}$\\
    Number of simulations & $N_\text{sim}$ & $>$\num{10000}\\
    Number of algorithm iterations & $N_\text{iter}$ & $K \cdot 4$\\
    Initialization vector & $\hat{\vect{\gamma}_\text{init}}$ & $\hat{\vect{\gamma}}_{\mathrm{init}}^{\mathrm{LMMSE}}$ (\ref{eq:lmmse})\\
    Unknown part of the \gls{csi} & $\lambda$ & \num{0.3}\\
    \bottomrule
    \end{tabular}
    \end{threeparttable}
    \end{table}

The default simulation configurations are summarized in Table~\ref{tab:initial-access-setup}. 
The device activity profile is generated randomly and independently for each device with a probability $\epsilon_a=0.1$, meaning that on average $\epsilon_a K = 50$ devices are active simultaneously. Or equivalently, the devices have an average duty cycle of 10\%, which is high for typical \gls{iot} applications~\cite{s21030913}.
The channel between the \gls{bs} and device $k$ is modelled as in (\ref{eq:prior_CSI_channel_model}).
The pilot sequence is randomly generated from a complex Gaussian distribution $s_k \cn{0}{1}$, and is assumed to be known by the \gls{bs}. Each device uses a pilot sequence of \num{10}~symbols. A random phase offset $\phi_k \sim\mathcal{U}_{[0, 2\pi]}$ is generated to simulate a \acrlong{cfo} (considered time-invariant over the preamble duration). The source code for all simulations can be accessed online\footnote{\url{https://github.com/wavecore-research/grant-free-random-access-partial-csi}}.

\subsection{Convergence of different initialization vectors}

\begin{figure}[tbp]
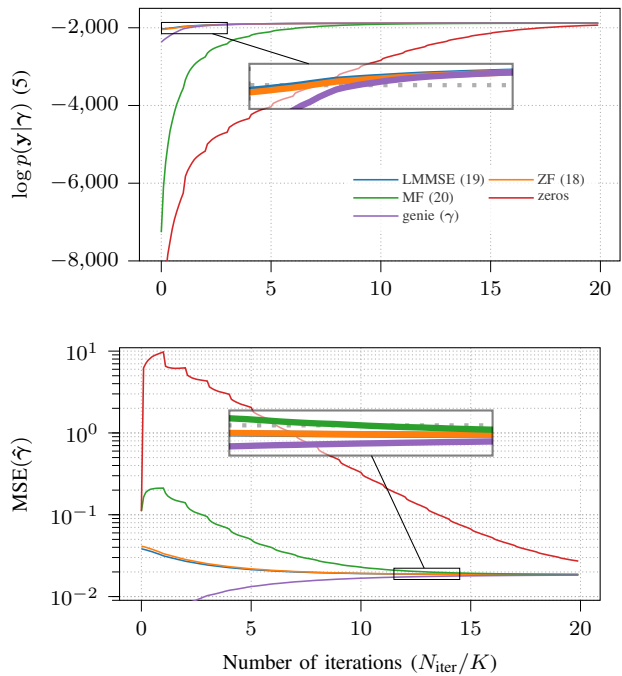

\centering
         \begin{subfigure}{0.9\columnwidth}
            \input{Fig/convergence_llh.tex}
        \end{subfigure}

         \begin{subfigure}{0.9\columnwidth}
            \input{Fig/convergence_mse.tex}
        \end{subfigure}
    \caption{The log-likelihood (\ref{eq:log_likelihood}) and \gls{mse} of the estimated activity indicators for different initialization vectors. While with all initialization vectors the genie-aided case is approximated, different number of iterations are required.}\label{fig:convergence}
\end{figure}

The convergence of different initializations is evaluated with respect to the genie-aided approach. In the genie-aided case, the algorithm is initialized with the real activity indicators, i.e., $\vect{\gamma}$. The convergence is assessed via the likelihood (\ref{eq:log_likelihood}) and the \gls{mse}. The former should monotonically increase with each iteration, while the \gls{mse} can vary as it can not directly be minimized. 
The performance of the different initialization vectors for $\hat{\vect{\gamma}}_{\text{init}}$ are depicted in Figure~\ref{fig:convergence}. The bottom figures zoom in on a smaller region to distinguish the performance of the initialization vectors when converging closer to the genie-aided case.  
While all initialization methods approximate the genie-aided case, the initialization vector has a non-negligible impact on the performance of the algorithm. An intuitive approach is to initialize with $\vect{0}$ because the activity probability is low and hence, on average, \SI{90}{\percent} of the devices are expected to be inactive. As illustrated in Figure~\ref{fig:convergence}, $\hat{\vect{\gamma}}_{\text{init}}=\vect{0}$ requires considerably more iterations to approach the other initialization methods.

\subsection{Impact of the quality of prior CSI}

\begin{figure}[tbp]
    \centering
        \input{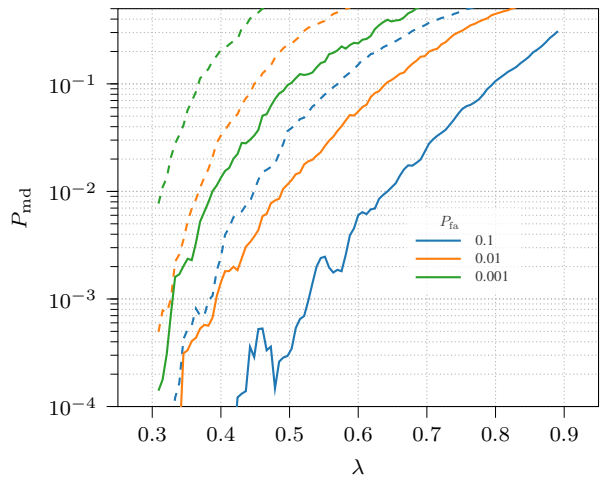}

    \definecolor{darkorange25512714}{RGB}{0,0,0}
    \definecolor{forestgreen4416044}{RGB}{0,0,0}
    \definecolor{lightgray204}{RGB}{0,0,0}
    \definecolor{steelblue31119180}{RGB}{0,0,0}
    
    \caption{Performance of the proposed algorithm (\ref{algo}) versus the \Gls{lmmse} estimator (\ref{rzf}) for different values of channel knowledge. The probability of miss detection is shown for different values of $\prob_{\mathrm{fa}}$ (\SI{10}{\percent}, \SI{1}{\percent}, \SI{0.1}{\percent}).}\label{fig:lambda_prob_gain}
\end{figure}

Figure~\ref{fig:lambda_prob_gain} illustrates the performance of the detector algorithms for different correlations between the actual channel and the known \gls{csi}, i.e., $\lambda$. 
With increased $\lambda$, and thus decreased channel knowledge, both the \gls{lmmse} estimator and the proposed algorithm have an increased probability of miss detection. The figure also demonstrates the gain of the proposed algorithm with respect to the \gls{lmmse} estimator. The algorithm outperforms the \gls{lmmse} estimator for all $\lambda$ and is most effective when the prior \gls{csi} has a strong correlation with the actual channel, and diminishes with decreased channel knowledge.

\subsection{Impact of the \acrlong{snr}}

\begin{figure}[tbp]
    \centering
        \input{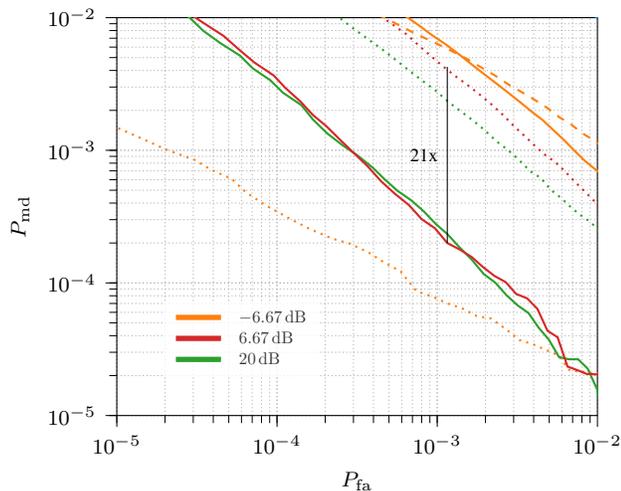}

    \definecolor{darkorange25512714}{RGB}{0,0,0}
    \definecolor{forestgreen4416044}{RGB}{0,0,0}
    \definecolor{lightgray204}{RGB}{0,0,0}
    \definecolor{steelblue31119180}{RGB}{0,0,0}
    \caption{The probability of false alarm and miss detection for different device \glspl{snr} for the \gls{lmmse} estimator when having full \gls{csi} ({\color{black} \ref{full}}) and partial \gls{csi} (\ref{rzf}), and the proposed iterative algorithm with partial \gls{csi} (\ref{algo}). No miss detection or false alarm occurred in the full \gls{csi} case for \gls{snr} values of \SI{-6.67}{\deci\bel} and \SI{20}{\deci\bel}.}\label{fig:fa_md_snrs}
\end{figure}

Figure~\ref{fig:fa_md_snrs} shows the false alarm and miss detection probability of the \gls{lmmse} estimator and the iterative maximum likelihood device activity detector for different device \glspl{snr}. The full \gls{csi} case is included as a baseline for comparison, where the full \gls{csi} is known instead of only a portion ($\lambda$). Fig.~\ref{fig:fa_md_snrs} demonstrates the large performance gain of the proposed algorithm with respect to the \gls{lmmse} estimator. The graph demonstrates that the iterative algorithm lowers the probability of miss detection by a factor of 21 for the same probability of false alarm.\footnote{Notably, the \gls{lmmse} estimator does not employ an iterative approach. Therefore, our proposed algorithm will be compared in future work with other iterative approaches.} The performance is only marginally increased for very low \glspl{snr} (below zero).

% \subsection{Achievable Rate}
% \begin{align}
%     R_k = \expt{\log_2(1+\text{SINR}_k)}
% \end{align}

% \begin{align}
%     R_{\text{sum}} = \sum\nolimits_{k \in \mathcal{K}_a} R_k
% \end{align}
% This figure of merits penalties when non-active users are included in the detector matrix, and also when active users were considered inactive. \gilles{Notably, the sum-rate could be higher even when not all active users are considered with respect to when all active users are included.}

\glsresetall
\section{Conclusion}\label{section:conclusion}
\glsresetall % reset glossaries for those reading only the conclusion

We formulated an iterative \gls{ml} algorithm to detect active devices using prior \gls{csi} when performing grant-free random access. Previous experimental work has demonstrated that, in many \gls{mmtc} applications, the \gls{csi} is less time-variant than assumed in theoretical models. Given the static nature of \gls{iot} devices, we have exploited this feature in the activity detection estimator. During grant-free access, the devices transmit a unique, but non-orthogonal preamble, which is used for activity detection. Next to this, the algorithm is also able to detect a device-specific phase offset, which could be caused by \gls{cfo}. The algorithm is numerically evaluated and compared to the conventional \gls{lmmse} estimator with full channel knowledge and partial \gls{csi}. The presented results indicate that the iterative algorithm converges and outperforms the conventional \gls{lmmse} estimator. For a \gls{snr} of \SI{6.67}{\deci\bel}, the probability of not detecting an active device is \num{21} times lower for the proposed iterative \gls{ml} estimator than the \gls{lmmse} estimator for the same probability of wrongly considering an inactive device as an active device.

This work can be extended to the cell-free or distributed case with geographically distributed access points. In that case, the partial \gls{csi} becomes access point-dependent.

%\input{Section/Appendix.tex}

%\section*{Acknowledgment}
%
%

\ifCLASSOPTIONcaptionsoff
  \newpage
\fi



\printbibliography

@book{kay1993fundamentals,
 author = {Nandi, Swagata and Kundu, Debasis},
 title = {Statistical Signal Processing},
 year = {2020},
 publisher = {Springer Singapore},
 doi = {10.1007/978-981-15-6280-8},
 source = {Crossref},
 url = {https://doi.org/10.1007/978-981-15-6280-8},
 subtitle = {Frequency Estimation},
 isbn = {9789811562792, 9789811562808},
}

@INPROCEEDINGS{Call2303:Grant,
AUTHOR="Gilles Callebaut and François Rottenberg and Liesbet {Van der Perre} and
Erik G. Larsson",
TITLE="{Grant-Free} Random Access of {IoT} devices in Massive {MIMO} with Partial
{CSI}",
BOOKTITLE="2023 IEEE Wireless Communications and Networking Conference (WCNC) (IEEE
WCNC 2023)",
ADDRESS="Glasgow, United Kingdom (Great Britain)",
DAYS="26",
MONTH=mar,
YEAR=2023
}

@inproceedings{GanesanSPAWC,
 author = {Ganesan, Unnikrishnan Kunnath and Bjornson, Emil and Larsson, Erik G.},
 title = {An Algorithm for Grant-Free Random Access in Cell-Free Massive {MIMO}},
 year = {2020},
 journal = {2020 IEEE 21st International Workshop on Signal Processing Advances in Wireless Communications (SPAWC)},
 volume = {},
 number = {},
 pages = {1--5},
 doi = {10.1109/spawc48557.2020.9154288},
 source = {Crossref},
 url = {https://doi.org/10.1109/spawc48557.2020.9154288},
 booktitle = {2020 IEEE 21st International Workshop on Signal Processing Advances in Wireless Communications (SPAWC)},
 publisher = {IEEE},
 month = may,
}

@article{SIG-093,
url = {http://dx.doi.org/10.1561/2000000093},
year = {2017},
volume = {11},
journal = {Foundations and Trends® in Signal Processing},
title = {Massive MIMO Networks: Spectral, Energy, and Hardware Efficiency},
doi = {10.1561/2000000093},
issn = {1932-8346},
number = {3-4},
pages = {154-655},
author = {Emil Björnson and Jakob Hoydis and Luca Sanguinetti}
}

@article{sherman1950adjustment,
  title={Adjustment of an inverse matrix corresponding to a change in one element of a given matrix},
  author={Sherman, Jack and Morrison, Winifred J},
  journal={The Annals of Mathematical Statistics},
  volume={21},
  number={1},
  pages={124--127},
  year={1950},
  publisher={JSTOR}
}

@article{GanesanTCOM,
 author = {Ganesan, Unnikrishnan Kunnath and Bjornson, Emil and Larsson, Erik G.},
 title = {Clustering-Based Activity Detection Algorithms for Grant-Free Random Access in Cell-Free Massive {MIMO}},
 year = {2021},
 journal = {IEEE Trans. Commun.},
 volume = {69},
 number = {11},
 pages = {7520--7530},
 doi = {10.1109/tcomm.2021.3102635},
 source = {Crossref},
 url = {https://doi.org/10.1109/tcomm.2021.3102635},
 publisher = {Institute of Electrical and Electronics Engineers (IEEE)},
 issn = {0090-6778, 1558-0857},
 month = nov,
}

@article{FenglerTIT,
 author = {Fengler, Alexander and Haghighatshoar, Saeid and Jung, Peter and Caire, Giuseppe},
 title = {Non-{Bayesian} Activity Detection, Large-Scale Fading Coefficient Estimation, and Unsourced Random Access With a Massive {MIMO} Receiver},
 year = {2021},
 journal = {IEEE Trans. Inf. Theory},
 volume = {67},
 number = {5},
 pages = {2925--2951},
 doi = {10.1109/tit.2021.3065291},
 source = {Crossref},
 url = {https://doi.org/10.1109/tit.2021.3065291},
 publisher = {Institute of Electrical and Electronics Engineers (IEEE)},
 issn = {0018-9448, 1557-9654},
 month = may,
}

@inproceedings{9049039,
 author = {Fengler, Alexander and Haghighatshoar, Saeid and Jung, Peter and Caire, Giuseppe},
 title = {Grant-Free Massive Random Access With a Massive {MIMO} Receiver},
 year = {2019},
 journal = {2019 53rd Asilomar Conference on Signals, Systems, and Computers},
 volume = {},
 number = {},
 pages = {23--30},
 doi = {10.1109/ieeeconf44664.2019.9049039},
 source = {Crossref},
 url = {https://doi.org/10.1109/ieeeconf44664.2019.9049039},
 booktitle = {2019 53rd Asilomar Conference on Signals, Systems, and Computers},
 publisher = {IEEE},
 month = nov,
}

@article{9091017,
 author = {Zhang, Qi and Jin, Shi and Zhu, Hongbo},
 title = {A Hybrid-Grant Random Access Scheme in Massive {MIMO} Systems for {IoT}},
 year = {2020},
 journal = {IEEE Open Access},
 volume = {8},
 number = {},
 pages = {88487--88497},
 doi = {10.1109/access.2020.2993597},
 source = {Crossref},
 url = {https://doi.org/10.1109/access.2020.2993597},
 publisher = {Institute of Electrical and Electronics Engineers (IEEE)},
 issn = {2169-3536},
}

@article{8456557,
 author = {Ding, Jie and Qu, Daiming and Jiang, Hao and Jiang, Tao},
 title = {Success Probability of Grant-Free Random Access With Massive {MIMO}},
 year = {2019},
 journal = {IEEE Internet of Things Journal},
 volume = {6},
 number = {1},
 pages = {506--516},
 doi = {10.1109/jiot.2018.2869003},
 source = {Crossref},
 url = {https://doi.org/10.1109/jiot.2018.2869003},
 publisher = {Institute of Electrical and Electronics Engineers (IEEE)},
 issn = {2327-4662, 2372-2541},
 month = feb,
}

@article{8454392,
 author = {Liu, Liang and Larsson, Erik G. and Yu, Wei and Popovski, Petar and Stefanovic, Cedomir and de Carvalho, Elisabeth},
 title = {Sparse Signal Processing for Grant-Free Massive Connectivity: {A} Future Paradigm for Random Access Protocols in the Internet of Things},
 year = {2018},
 journal = {IEEE Signal Process Mag.},
 volume = {35},
 number = {5},
 pages = {88--99},
 doi = {10.1109/msp.2018.2844952},
 source = {Crossref},
 url = {https://doi.org/10.1109/msp.2018.2844952},
 publisher = {Institute of Electrical and Electronics Engineers (IEEE)},
 issn = {1053-5888, 1558-0792},
 month = sep,
}

@article{callebaut2021grant,
    title        = {{Grant-Free Random Access in Massive MIMO for Static Low-Power IoT Nodes}},
    author       = {Callebaut, Gilles and Van der Perre, Liesbet and Rottenberg, Fran{\c{c}}ois},
    year         = 2021,
    booktitle    = {2022 42nd Symposium on Information Theory and Signal Processing in the Benelux},
    pages        = {46--51},
    url          = {arXiv preprint arXiv:2110.07927}
}

@article{s21030913,
    title        = {{The Art of Designing Remote IoT Devices—Technologies and Strategies for a Long Battery Life}},
    author       = {Callebaut, Gilles and Leenders, Guus and Van Mulders, Jarne and Ottoy, Geoffrey and De Strycker, Lieven and Van der Perre, Liesbet},
    year         = 2021,
    journal      = {Sensors},
    volume       = 21,
    number       = 3,
    doi          = {10.3390/s21030913},
    issn         = {1424-8220},
    url          = {https://www.mdpi.com/1424-8220/21/3/913},
    article-number = 913,
    pubmedid     = 33572897
}

@book{marzetta2016fundamentals,
    title        = {Fundamentals of {M}assive {MIMO}},
    author       = {T. L. Marzetta and E. G. Larsson and H. Yang and  H. Q. Ngo},
    year         = 2016,
    publisher    = {Cambridge University Press}
}

@article{cell-free,
    title        = {Cell-Free Massive {MIMO} Versus Small Cells},
    author       = {Ngo, Hien Quoc and Ashikhmin, Alexei and Yang, Hong and Larsson, Erik G. and Marzetta, Thomas L.},
    year         = 2017,
    journal      = {IEEE Trans. Wireless Commun.},
    volume       = 16,
    number       = 3,
    pages        = {1834--1850},
    doi          = {10.1109/TWC.2017.2655515}
}

@INPROCEEDINGS{9443344,
  author={Wielandt, Stijn and Dafflon, Baptiste},
  booktitle={2020 54th Asilomar Conference on Signals, Systems, and Computers}, 
  title={A Local LoRa Based Network Protocol with Low Power Redundant Base Stations Enabling Remote Environmental Monitoring}, 
  year={2020},
  volume={},
  number={},
  pages={520-523},
  doi={10.1109/IEEECONF51394.2020.9443344}}

\end{document}